\documentclass{iopart}
\usepackage{epsfig}

\begin{document}

\def\bq{\begin{eqnarray}}
\def\eq{\end{eqnarray}}
\def\l{\langle}
\def\r{\rangle}

\jl{4}

\title[QCD sum rules on the light cone]{QCD sum rules on the light cone and
$B \rightarrow \pi$ form factors}


\author{S Weinzierl\dag\footnote[1]{Talk given by S Weinzierl at the
    UK Phenomenology Workshop on Collider Physics, Durham, 19-24 September 1999} and O Yakovlev\ddag}
\address{\dag NIKHEF, P.O. Box 41882, 1009 - DB Amsterdam, The Netherlands}
\address{\ddag Randall Laboratory of Physics, University of Michigan, Ann Arbor, Michigan 48109 - 1120, USA}


\begin{abstract}
The semileptonic decay $B \rightarrow \pi \bar{l} \nu$ is one of the most important
reactions for the determination of the CKM matrix element $|V_{ub}|$.
However, in order to extract $|V_{ub}|$ from data one needs an accurate
theoretical calculation of the hadronic matrix element describing the
$B$ to $\pi$ transition.
QCD sum rules, based on operator-product expansion on the light-cone,
provide a reliable approach to this aim.
QCD corrections and higher twist contributions can be taken systematically
into account. 
\end{abstract}

\section{Motivation: CP violation and the CKM matrix}

Within the Standard model CP violation is parameterized by a complex phase
in the CKM-matrix $V_{CKM}$. The various entries of $V_{CKM}$ are known
with different accuracy: The best known matrix elements are $V_{ud}$ and
$V_{us}$, the first one is obtained from the comparison of
super-allowed nuclear $\beta$-decay with $\mu$-decay, the latter one
from the decay $K \rightarrow \pi \bar{l} \nu$.
Among the matrix elements which are fairly well known are $V_{cd}$ (obtained
from single charm production in deep-inelastic $\nu N$-scattering and from
semileptonic decays of charmed mesons), 
$V_{cs}$ (which can be obtained from the decay $D \rightarrow \bar{K} \bar{l} \nu$) and $V_{cb}$. 
The latter one can be measured either inclusively, using
$B \rightarrow D$ transitions or exclusively in the semileptonic 
$B \rightarrow D$ transition.
Among the matrix elements which are least well known are $V_{ub}$, $V_{td}$,
$V_{ts}$ and $V_{tb}$. The matrix elements involving the top quark
may be obtained from $B_d$-$\bar{B}_d$ mixing ($V_{td}$),
$B_s$-$\bar{B}_s$ mixing ($V_{ts}$) and single top production
($V_{tb}$).
The remaining one, $V_{ub}$, can be obtained either
from the lepton spectrum in inclusive $B \rightarrow X \bar{l} \nu$ decays
or from exclusive semileptonic $B$-decays.

In order to extract the entries of the $V_{CKM}$ matrix from the experimental data
one needs certain input information from theory. In many cases the theoretical
calculations have to rely on non-perturbative methods for QCD.
Among the techniques which have been used are: Chiral perturbation theory (for the
extraction of $V_{us}$), heavy quark effective theory (for $V_{cb}$), lattice
calculations (for $V_{ub}$, $V_{cs}$ and $V_{td}$), QCD sum rules (for
$V_{ub}$ and $V_{cs}$ 
)
as well as quark models (for $V_{ub}$ and $V_{cs}$).
Among these the first four enjoy the property that they are based on first principles, whereas
quark models are, as the name already indicates, phenomenological models, whose systematic
errors are difficult to quantify. 

In this talk we will focus on the extraction of $|V_{ub}|$ from the exclusive
semileptonic decay $B \rightarrow \pi \bar{l} \nu$, using QCD sum rules
techniques.
In the next section we explain the basic principles underlying 
the sum rule technique.
QCD sum rules are based on the operator product expansion (OPE). We 
mention briefly the differences between a short-distance
expansion and an expansion on the light-cone.
In the last section we will focus on the sum rule for the decay
$B \rightarrow \pi \bar{l} \nu$ and give numerical results.

\section{The QCD sum rule technique}

The derivation of a QCD sum rule \cite{Shifman:1979bx} involves the following six steps:

Step 1: Take a correlation function. For example, in order to calculate the B-meson
decay constant $f_B$ using QCD sum rules, one would start from the correlation function
\bq
\Pi(q^2) & = & i \int d^4x e^{iqx} \l 0 | T \left( \bar{u}(x) i \gamma_5 b(x), 
 \bar{b}(0) i \gamma_5 u(0) \right) | 0 \r.
\eq

Step 2: Write a dispersion relation for $\Pi(q^2)$:
\bq
\Pi(q^2) & = & \frac{1}{\pi} \int ds \frac{\mbox{Im} \Pi(s)}{s-q^2} \; + \; \mbox{subtractions}
\eq

Step 3: Express the absorbative part of $\Pi(q^2)$ in terms of hadronic quantities.
In our example above this would involve $f_B$.

Step 4: Calculate the correlation function in the asymptotic region using QCD
and operator product expansion.

Step 5: Use quark-hadron duality to relate the hadronic representation from step 3
to the absorbative part of the QCD calculation from step 4.

Step 6: The sum rule can be improved  by applying a Borel transformation
to both sides.

These steps provide a ``cooking recipe'' for QCD sum rules. A few comments:
The QCD calculation in step 4 is based on the operator product expansion.
A nonlocal composite operator like $j(x) j(0)$ (in the example above
we have $j(x) = \bar{u}(x) i \gamma_5 b(x)$) is expanded into a series
of well-defined local operators ${\cal O}_n$:
\bq
j(x) j(0) & = & \sum\limits_n C_n {\cal O}_n
\eq
This separates soft and hard physics: The Wilson coefficients $C_n$ contain the
information on short distance physics and can be calculated using perturbation
theory. The matrix elements of the local operators ${\cal O}_n$ parameterize
the long distance physics. They are universal non-perturbative quantities.
There are two version of the operator product expansion: The short distance
expansion and the expansion on the light-cone. In the former one the various
local operators are classified by their dimensions, where as in the latter
one the classification goes by twist (dimension minus spin).

The sum rule can be improved by applying the Borel operator to both
the hadronic representation and the QCD calculation. The Borel
operator is given by
\bq
\hat{B} & = & \lim\limits_{Q^2 \rightarrow \infty, n \rightarrow \infty, Q^2/n = M^2} \frac{1}{(n-1)!} \left( Q^2 \right)^n \left( - \frac{d}{dQ^2} \right)^n.
\eq
The Borel transformation gives rise to an exponential supression of the higher resonances
and the continuum in the hadronic representation.
The power corrections in the QCD calculations are supressed by factors of $(1/M^2)^n$.
An upper limit on $M^2$ is obtained by requiring that the contributions from the
higher resonances and the continuum should not be too large.
A lower limit on $M^2$ is obtained from requiring that terms supressed by
powers of $1/M^2$
should be subdominant.
It is important to check that this defines a window, in which the final results
are insensitive to the variation of the Borel parameter $M^2$.

\section{$B \rightarrow \pi \bar{l} \nu$ and sum rules on the light cone}

The relevant hadronic matrix element is parameterized by two form factors $f^+$ and $f^-$ :
\bq
\l \pi(q) | \bar{u} \gamma_\mu b | B(p+q) \r & = & 2 f^+(p^2) q_\mu
+ \left( f^+(p^2) + f^-(p^2) \right) p_\mu
\eq
If we neglect lepton masses only the form factor $f^+$ gives a contribution to the decay width.
In order to obtain $f^+$ from QCD sum rules we start from the correlation function
\bq
F_\mu(p,q) & = & i \int d^4x e^{ipx} \l \pi(q) | T \left(
 \bar{u}(x) \gamma_\mu b(x), \bar{b}(0) i m_b \gamma_5 d(0) \right) | 0 \r \nonumber \\
& = & F(p^2,(p+q)^2) q_\mu + \tilde{F}(p^2,(p+q)^2) p_\mu .
\eq
It can be shown \cite{Belyaev:1995zk} that a short distance expansion is only useful in the soft
pion limit ($q \rightarrow 0$). A better approach is provided by the expansion
around the light cone $x^2=0$ with operators of increasing twist.
In our case the leading twist-2 contribution is given by
\bq
\l \pi(q) | \bar{u}(x) \gamma_\mu \gamma_5 d(0) | 0 \r & = & 
-i q_\mu f_\pi \int\limits_0^1 du e^{iuqx} \varphi_\pi(u) .
\eq
$\varphi_\pi(u)$ is known as the twist-2 pion light-cone wave function. 
The correlation function $F$ can now be written as a convolution of the
pion wave function with a hard scattering amplitude $T(p^2,(p+q)^2,u)$:
\bq
F(p^2,(p+q)^2) & = & - f_\pi \int\limits_0^1 du \varphi_\pi(u) T(p^2,(p+q)^2,u)
\eq
The hard scattering amplitude $T$ can be calculated within perturbation theory,
whereas the light-cone wave function contains the non-perturbative information.
At NLO both the hard scattering amplitude $T$ and the wave function $\varphi_\pi$
depend on a factorization scale $\mu$. The evolution of $\varphi_\pi$ with this
scale can be calculated perturbatively \cite{Kadantseva:1986kb} . The
Brodsky-Lepage evolution kernel for the pion wave function
is analog to the Altarelli-Parisi kernel describing the evolution of
parton densities.
It turns out that it is convenient to express the wave function in terms
of Gegenbauer polynomials. To leading order these Gegenbauer
polynomials are eigenfunctions of the evolution kernel.
However this is no longer true at next-to-leading order and 
mixing effects have to be taken into account.
At present the twist 2 contributions have been calculated to next-to-leading order \cite{Khodjamirian:1997ub,Bagan:1997bp,Khodjamirian:1999hb}, 
and twist 3 and twist 4 to leading order \cite{Belyaev:1993wp}.
Fig. 1 shows the result for the $B \rightarrow \pi$ form factor obtained
from the light-cone sum rule in comparison to lattice results.
\begin{figure}
\centerline{
\epsfig{file=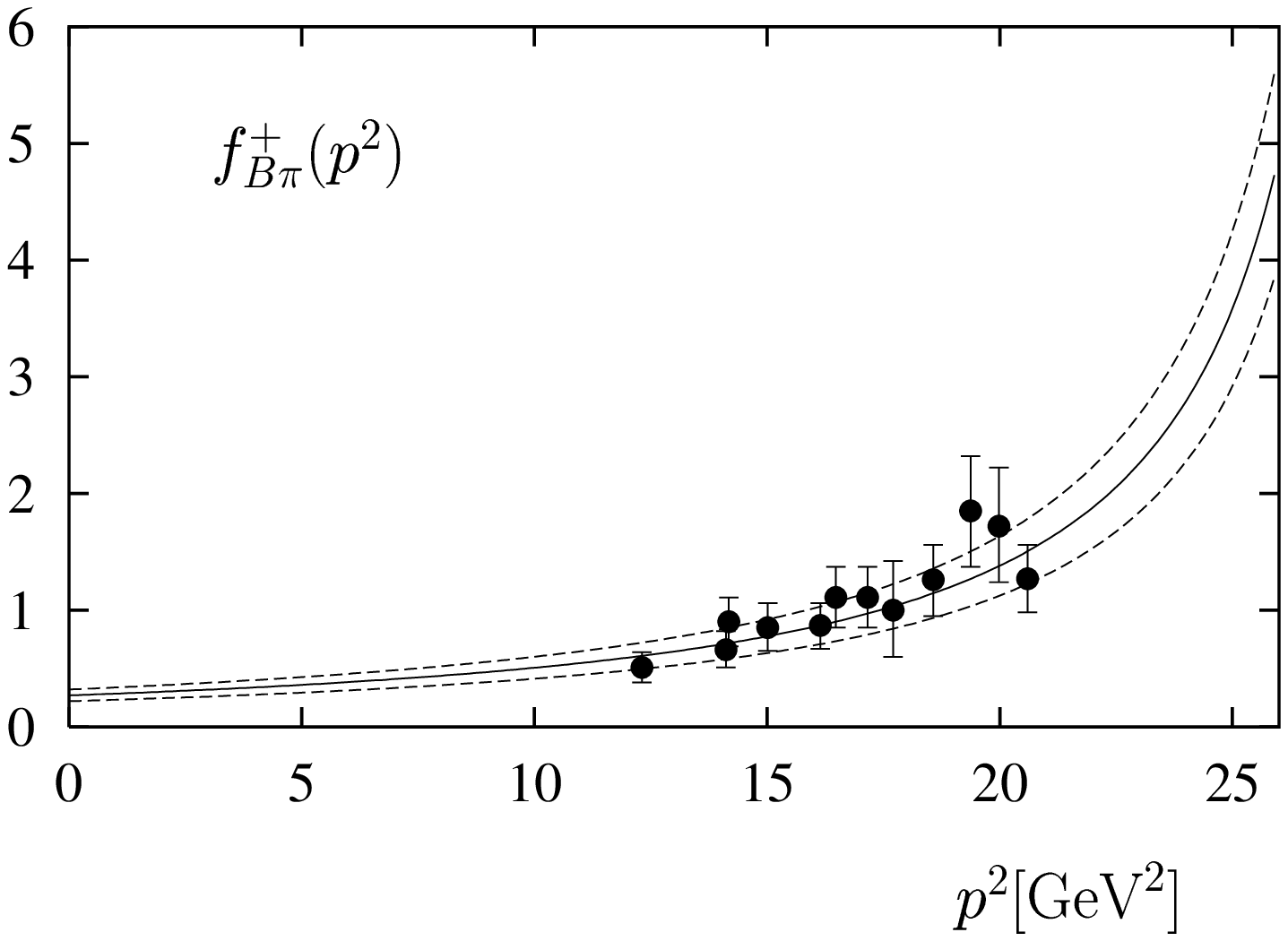,
bbllx=10pt, bblly=230pt, bburx=495pt, bbury=505pt,
width=8cm
}
}
\caption{\label{fig1} \it The LCSR prediction \cite{we} for the $B \rightarrow \pi$ form factor
$f^+_{B\pi}(p^2)$ (solid curve) in comparison to lattice results 
\cite{Flynn:1996rc}.
The estimated theoretical uncertainties are shown by dashed curves.}
\end{figure}
The theoretical error is estimated from a variation of the Borel parameter,
the $b$-quark mass, the threshold parameter $s_0$, the quark condensate 
density and the normalization scale. Furthermore our insufficient
knowledge of the shape of the light-cone wave functions, unknown higher order
perturbative corrections or higher twist effects contribute
as well to the theoretical error.
Our final result is fitted to the parameterization \cite{Becirevic:1999kt}
\bq
f^+(p^2) & = & \frac{f^+(0)}{\left( 1 -  \frac{p^2}{m_{B^*}^2} \right)
  \left( 1 - \alpha \frac{p^2}{m_{B^*}^2} \right)} 
\eq
with
\bq
f^+(0) = 0.27 \pm 0.05, & & \alpha = 0.35 \pm 0.01.
\eq
The form factor $f^+(p^2)$ enters the partial decay width
\bq
\frac{d \Gamma}{dp^2} & = & \frac{G_F^2 |V_{ub}|^2}{24 \pi^3} 
\left( E_\pi^2 - m_\pi^2 \right)^{3/2} \left( f^+(p^2) \right)^2.
\eq
Integration yields
\bq
\Gamma(B^0 \rightarrow \pi^- e^+ \nu_e) & = & (6.7 \pm 2.8) |V_{ub}|^2 
\mbox{ps}^{-1}.
\eq
Comparing this with the experimental value \cite{Alexander:1996qu}
\bq
\Gamma(B^0 \rightarrow \pi^- e^+ \nu_e) & = & (1.15 \pm 0.34) \cdot 10^{-4} 
\mbox{ps}^{-1}
\eq
one obtains \cite{we}
\bq
|V_{ub}| & = & 0.0041 \pm 0.0005 \pm 0.0006,
\eq
where the first error corresponds to the current experimental uncertainty
and the second error to the estimated theoretical uncertainty.
One may expect more precise data from the forecoming experiments of Babar
and Belle. On the theoretical side, an improved knowledge of the pion 
light-cone wave function and perturbative corrections to twist 3
would reduce the theoretical error.

\section*{References}


\end{document}